\documentclass[
nofootinbib,
superscriptaddress,
pra,
amsmath,
amssymb,
notitlepage,
floatfix,
]{revtex4-2}

\usepackage[utf8]{inputenc}
\usepackage{geometry}
\usepackage{subcaption}
\usepackage{xcolor}
 \usepackage{graphicx}

\begin{document}
\preprint{APS/123-QED}

\title{Elastic wave control in reticulated plates using Schwarz primitive cells}

\author{Aida Hejazi Nooghabi}
\affiliation{Department of Civil, Environmental and Geomatic Engineering, ETH Zurich, Zurich, Switzerland}

\author{Henrik R. Thomsen}
\affiliation{Institute for Geophysics,  ETH Zurich, Zurich, Switzerland}

\author{Bao Zhao}
\affiliation{Department of Civil, Environmental and Geomatic Engineering, ETH Zurich, Zurich, Switzerland}

\author{Andrea Colombi}
\affiliation{Zurich University of Applied Sciences (ZHAW), Winterthur, Switzerland}

\maketitle
\date{}
\section{Abstract}

In this work the Schwarz primitive unit cell is used as the building block of different types of metastructures for steering and focusing elastic vibrations. The emergence of a Bragg-type bandgap when constructing a 2D plate from such unit cells is experimentally validated. It is demonstrated that both increasing the mass and the porosity of the Schwarz primitive lead to a decrease in the frequency of the out-of-plane propagating wave targeted in this study.
By arranging these modified Schwarz primitive unit cells in constant and graded layouts, 2D plates with an embedded metabarrier and a metalens are numerically designed. 
The metabarrier protects an interior area of the plate from the propagating waves on a wide frequency band ($\sim$1.4 to $\sim$3.4 kHz). Equally, the refractive index profile necessary for gradient index lenses is  obtained via a progressive  variation of the added mass or, alternatively, the porosity of the unit cell over a rectangular area. For the first time, bending of the out-of-plane mode towards the focusing point is practically validated in a challenging mesoscale experiment requiring the assembly of different 3D printed sections of the plate. The increased porosity design is advantageous not only in terms of overall lightweight, but also towards additive manufacturing as it requires less material.
\\

keywords: Schwarz primitive cell, elastic wave propagation, bandgap, metabarrier, Luneburg lens

\section{Introduction}
 
 Metamaterials - \textcolor{black}{with their engineered structures} - have been heavily exploited in the past couple of decades both for static and dynamic applications. They have proven useful for a variety of applications such as sound absorption \citep{acoustic-absorption}, elastic wave guiding \citep{park-guiding}, energy harvesting \citep{CHEN-harvesting-review,Zhao-harvesting}, mechanical neural smart devices \citep{dub2023binary}, and seismic and vibration shielding \citep{Miniaci-seismic-shielding,zhao2023nonlinear,kyriakosPhysRevApplied}, to name a few.
One of the most important features of such structures is the so-called bandgap \citep{brillouin}, defined as the frequency range where the propagation of waves is prohibited \citep{brillouin}. The well-known mechanisms behind this feature are either Bragg scattering \citep{kittel2004introduction}, local resonance \citep{Colombi2016forests,ZHOU2012041001}  or the coalescence of the two \citep{cenedese-coale,KRUSHYNSKA201730,lee-resonance-bragg,Aguzzi2022}. While Bragg scattering emerges due to the periodicity in the arrangement of the unit cells \citep{brillouin} constituting the metamaterial structure, the local resonance mechanism relies on the resonance of the unit cell and is independent upon its arrangement. Local resonances could appear in both periodic and non-periodic distributions of such cells \citep{rupin2014experimental}. 
The opening frequency of the bandgap due to Bragg scattering in phononic crystals is determined by the relationship between the wavelength of the propagating wave and the spacing between unit cells. Contrary, in locally resonant metamaterials,  the  opening frequency is only dependent on the resonant frequency of a single resonator. Additionally, the frequency range of the bandgap in phononic crystals is often wider than the one in resonant metamaterials. These complementary characteristics have led to studies where a coalescence of the Bragg scattering and local resonance has been implemented to obtain tunable wide bandgaps \citep{Aguzzi2022}.

In acoustics, metamaterials have been widely used for noise absorption and wave guiding, whereas, in elasticity a broader range of applications such as guiding, focusing, and energy harvesting have been explored. 
More specifically, elastic wave focusing based on GRadient INdex (GRIN) lenses has been used in a variety of structures and gradient profiles. For instance, Maxwell and Luneburg lenses have been numerically tested based on lattice structures in 2D by leveraging both Bragg and local resonance mechanisms in order to focus the out-of-plane waves in the kHz regime \citep{Aguzzi2022}. A GRIN metasurface for microwave imaging has been numerically and experimentally tested in 1D and 2D  at the GHz regime \citep{Datta-microwave} and at kHz regime, focusing of surface acoustic waves has been experiementally achieved based on resonant metasurfaces \citep{SAW}.
A wider variety of GRIN lenses including not only Maxwell and Luneburg, but also Eaton and 90$^{\circ}$ rotating lenses have been built based on the variation in the thickness of the plates for the flexural waves \citep{climente-gradient}.

In this work, we investigate how the Schwarz primitive cell \citep{schwarz1972gesammelte} which belongs to the family of Triply Periodic Minimal Surfaces (TPMS) can be leveraged in the design of meta structures. 
The Schwarz primitive is symmetric in 3D and is known to show significant strength to compression and fracture toughness, therefore making it a candidate with high potentials for scaffolds in biomimetic designs and TPMS-based scaffolds \citep{MONTAZERIAN201798, biomimetic, pan2020}.
TPMS structures can offer enhanced mechanical properties compared to truss-like structures. The continuous surface geometry of TPMS structures and gradual change in curvature allows for an efficient load distribution and stress transfer across the surface. This minimizes stress concentrations, improving stress distribution and eventually makes the frame less prone to fatigue damages \citep{SOKOLLU2022103199, pan2020}. 
\textcolor{black}{The Schwarz primitive cell has previously been investigated for its buckling characteristics \citep{VIET}, numerically and experimentally validated for topological elastic wave guiding \citep{guo2022minimal}, and has also been explored for vibration mitigation purposes in the realm of acoustic black holes \citep{ma2023semi}.
An in-depth investigation on the relationship between the volume ratio and bandgap behavior \citep{silva2023}  as well as the acoustic properties of Schwarz primitive cell \citep{Lu2023} and mechanical vibration bandgaps \citep{ELMADIH} have  been performed. Despite such different types of investigations on the dynamic properties of the Schwarz primitive cell in the context of acoustic and elastic metamaterials [29–31], their potential in the realm of focusing and mitigation of elastic waves has not been thoroughly studied. }

Interestingly, biological arrangements with geometries similar to TPMS have been observed \citep{Min-seaurchin}. For instance, a slice of a sea urchin skeletal plate displays similarities to the Schwarz primitive geometry \citep{Alketan-review, Min-seaurchin}.  
Also, the structure of amphiphilic membranes that separate oil and water, also known as the “plumber's nightmare”, form arrangements that resemble the Schwarz primitive surface \citep{GANDY2000579}.
Despite different types of investigations on the dynamic properties of TPMS (more precisely the Schwarz primitive cell) in the context of elastic metamaterials \citep{ABUEIDDA201820, guo2022minimal, ma2023semi}, their potential in the context of focusing and mitigation has not been thoroughly studied. 

%

In this paper, we first study the dynamic behaviour of a Schwarz primitive cell by extracting the dispersion relation of a unit cell made of versatile plastic. We validate the presence of a bandgap both numerically and experimentally. We next build plate-like metastructure by locally varying the geometry in the medium to  manipulate elastic waves for vibration mitigation (metabarriers) and focusing (GRIN lenses) purposes. \textcolor{black}{The particle vibration of interest is dominated by out-of-plane motion}, which is analogous to the A$_0$ mode of Lamb waves propagating in thin plates \citep{Royer1980ElasticWI}.
 Through large scale, parallel finite element time domain simulations 
 we numerically validate the performance of such devices based on adding mass and increasing porosity in the unit cell. 
 We find that the increased porosity not only adds to the metastructure's lightweight characteristics, but also makes it very suitable for 3D printing (less material) and hands-on experimentation. The concept of increasing porosity as a design variable for a GRIN lens is additionally verified through laboratory experiments conducted on a 3D printed prototype. Our findings on the functionality of such a metastructure pave the way for further applications of bio-mimetic structures in relation to wave propagation. 

\section{Dispersion relation of the Schwarz primitive cell}
\subsection{Unit cell design}
The geometry of the TPMS structures can be defined by trigonometric functions and in the case of the Schwarz primitive cell, the 3D surface is obtained by solving the following implicit equation:

\begin{equation}
    cos(X)+cos(Y)+cos(Z) = C
    \label{eq:sch-P-formula}
\end{equation}
where $X=2\pi x/L_x$, $Y=2\pi y/L_y$, $Z=2\pi z/L_z$, $L_x$, $L_y$ and $L_z$ are the size of the unit cell in the $x$, $y$, and $z$ directions and $C$ is a constant \citep{schwarz1972gesammelte, pan2020}.
In this study, the size of the unit cell along the $x$, $y$ and $z$ directions is 3 cm and the value of $C$ is zero. The resulting geometry is generated in MATLAB (See Figure. \ref{fig1-num}a). By adding an offset, the generated surface is thickened by 0.5 mm and the obtained geometry is then exported in STL format. \textcolor{black}{The selected dimensions and thickness of the unit cell ensure the wave propagation properties required for designing metastructures, making it optimal for both experimental and numerical analyses.} For the following time domain simulations, the full geometry is processed in Coreform Cubit, where the solid hexahedral mesh  elements are generated.  
Once the geometry is built, thickened, healed, and meshed, the dispersion relation of the cell can be numerically extracted. In the following, we show the dynamic behaviour of the Schwarz primitive unit cell through dispersion relations obtained both numerically and experimentally. 

\subsection{Numerical validation of the dispersion relation}
\begin{figure}[h]
\centering
\includegraphics[trim={0 190 5 100},clip, keepaspectratio,scale=0.85]{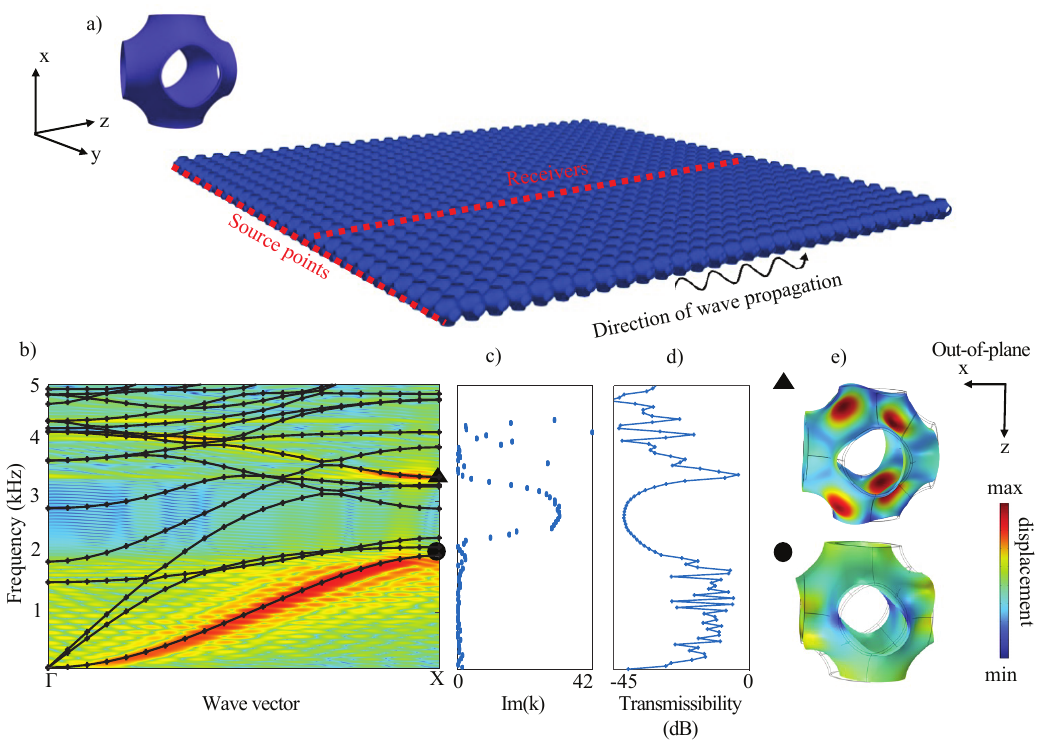}
\caption{(a) The Schwarz Primitive unit cell with a length of 3 cm along the $x$, $y$ and $z$ directions and the configuration of the 2D plate used in time domain simulations built by tessellating the unit cell in two directions. (b) Dispersion relation of the unit cell along the $\Gamma$-$X$ direction obtained via Bloch analysis in COMSOL (black curves) and the outcome of 2D Fourier transform applied to time domain data (colored background plotted in dB). (c) Imaginary part of the wavenumber obtained from the time domain data. (d) Transmissibility retrieved from time domain data. (e) Mode shapes at the opening and closing frequencies of the bandgap. }
\label{fig1-num}
\end{figure}

Through simulations, we obtain the dispersion relation i.e., the relationship between the frequency $f$ and real part of the wave number $k$ based on two different methods, namely as eigenfrequency analysis  \citep{comsol_manual} and 2D FFT method \citep{2d_fft}. 
The eigenfrequency analysis is performed using the software COMSOL Multiphysics 6.1, where the dispersion relation of the unit cell discretized into solid tetrahedral mesh elements is solved through the finite element method. This analysis is based on the Bloch theorem, where Bloch periodic boundary conditions are applied to the unit cell and the dispersion relation $\omega = \omega(k)$ is solved by varying the wave vector $k$ along direction of the wave propagation \citep{bloch1929quantenmechanik}. We choose a conventional cubic unit cell and focus only on the wave propagation along $\Gamma$-$X$.  
As the out-of-plane mode, propagating along the $z$-direction, is of interest, we retrieve and show the propagating modes only along this direction, however, in COMSOL all the modes (including in-plane and out-of plane) are generated (See Figure. \ref{fig1-num}b).

In order to correctly identify which modes are excited by an out-of-plane source, we next extract the dispersion relation along the $z$-direction through time domain simulations. The finite reticulated plate is obtained by tessellating 30 unit cells along both the $y$ and $z$ directions. Solid hexahedral mesh elements are generated through the Python APIs of the Coreform Cubit software (See Figure. \ref{fig1-num}a). In this set of computations, the material properties were set to:  Young's  modulus $E$ = 1.0 GPa, Poisson's ratio $\nu$ = 0.39 and mass density  $\rho$ = 930 kg/m$^3$. 
The model shown here consists of approx $10^6$ hex elements. Time domain simulations are all performed in SPECFEM3D, an open-source code that implements a highly parallelised spectral element method for solving elastodynamic equations \citep{Komatitsch1999}.
Since we are interested in the out-of-plane mode, only the polarization along the $x$-axis is excited at points marked in Figure.\ref{fig1-num}a. The dispersion relation for the given mode can be obtained by extracting the displacement field along the propagation direction and applying a 2D Fourier transform to the data (See Figure.~\ref{fig1-num}b). The dark orange features show the propagating out-of-plane modes.
Both the Bloch and time-domain analyses reveal a bandgap between 2.021 and 3.437 kHz for the A$_0$ mode. The results of both methods displayed in Fig. \ref{fig1-num}b suggest a Bragg-type mechanism for the observed bandgap of the out-of-plane modes.
The latter can be validated via recovering the imaginary part of the wavenumber using the IWC method proposed by \citep{VanDamme-iwc}. A behaviour typical of the Bragg bandgap is revealed (See Figure. \ref{fig1-num}c).
The transmissibilty shown in Figure.~\ref{fig1-num}d also indicates the presence of the bandgap already identified in Figure.~\ref{fig1-num}b.
Mode shapes at the opening and closing frequencies of the bandgap  (indicated by a circle and triangle in Figure. \ref{fig1-num}b) are shown in Figure. \ref{fig1-num}e.
While the opening mode of the bandgap is a purely out-of-plane mode, the closing mode includes a  mixture of the out-of-plane and a torsional mode along the $z$ direction.

\subsection{Experimental validation of the \textcolor{black}{bandgap}}

\begin{figure}[t]
\centering
\includegraphics[trim={5 180 5 160},clip, keepaspectratio,scale=0.8]{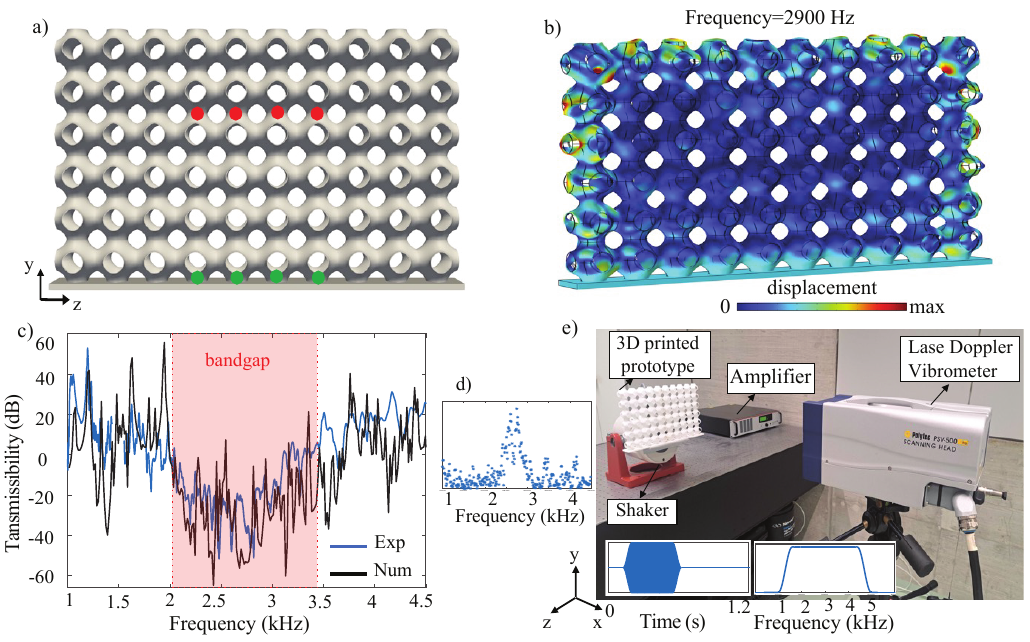}

\caption{(a) Finite plate for frequency domain simulations built by replicating the unit cell in 2D. The green and red points mark the positions where data were extracted to calculate the transmissibility. (b) Mode shape of the finite structure at frequency 2900 Hz. Although the frequency lies within the bandgap, some modes appear in the boundary cells. (c) Transmissibility obtained from the  numerical (black) and experimental (blue) data. The pink shaded regions marks the range of the expected bandgap, obtained from the infinite model. (d) Imaginary part of the wave-number based on experimental data. (e) Experimental set-up,  along with the signal used as the source wavelet in the experiment. }\label{fig2-exp}
\end{figure}

The versatility of additive manufacturing, also for thin walled structures like this one, allows to validate the bandgap of the Schwarz primitive cell experimentally.
In order to determine the size of the plate (i.e., the number of cells along the two directions) that allows the wavefield to be identified but also satisfies printability constraints \textcolor{black}{in terms of size}, we must perform a numerical simulation on a finite structure in the frequency domain using COMSOL Multiphysics 6.1. A plate consisting of 10 by 6 unit cells along the $y$ and $z$ directions, respectively mounted on top of a thin support plate of 4 mm thick necessary to attach the structure to an electrodynamic shaker, provided promising results by considering the elastic parameters defined earlier and an isotropic loss factor of $\eta = 0.002$. 
The structure is excited at discrete frequencies between 1 and 5 kHz with a constant displacement (\textbf{\textit{u}}) along the $x$ direction through the supporting plate (\textbf{\textit{u}}=($u_{x}$,0,0)).
The transmissibility is then computed in dB:

\begin{equation}
    Transmissibility = 20*log \left( \frac{u_{top}(f)}{u_{bottom}(f)} \right),
    \label{eq:trans}
\end{equation} 
with $u_{bottom}$ and $u_{top}$ defined as the out-of-plane component of displacement extracted at specific points in the model. The numerical study shows that close to the borders, the wavefield experiences boundary effects from the free sides and the bottom plate. Figure. \ref{fig2-exp}b illustrates an example of such existing modes at a frequency of 2900 Hz,  which lies within the bandgap. Although no propagation in the structure is expected at this frequency, the displacement at the borders of the modelled prototype does not completely vanish. To mitigate the effects of this boundary noise, $u_{top}$ is selected after the fourth row and not at the last row. Additionally, both the $u_{bottom}$ and $u_{top}$ are averaged over several points (See Figure. \ref{fig2-exp}a). 

The numerical transfer function is shown in Figure. \ref{fig2-exp}c. The pink region specifies the bandgap range obtained through eigenfrequency analysis on an infinite model calculated previously. 
We observe a very good match between the two indicating that 10 by 6 unit cells guarantee the expected functionality. Therefore, we proceed with a prototype of identical size 3D printed via Selective Laser Sintering (SLS). Similar to the numerical case, a thin plate of 30.5 by 3 cm and a thickness of 4 mm was added to the bottom of the design to act as an excitation base. Due to limits in the maximum printable size along the $z$-direction imposed by the manufacturer, the prototype was assembled out of two smaller samples each consisting of 5 by 6 unit cells (along the $y$ and $z$ directions, respectively).
The structure was then attached to a shaker using super glue. The A$_0$ mode is excited in the structure by sending a sweep signal with the duration of 1.2 seconds carrying frequencies between 1 and 5 kHz as shown in the bottom panels of Figure. \ref{fig2-exp}e. As seen from the experimental set-up in Figure. \ref{fig2-exp}e, we measure the out-of-plane velocity of the excited propagating mode using a one-component Scanner Laser Doppler Vibrometer (SLDV) Polytec PSV500. The experimental transmissibility of the prototype structure is determined the same way as in the numerical tests. In this case, we take the ratio of the velocity between the measurements at the top and the bottom of the structure. The results are shown in Figure. \ref{fig2-exp}c. 
 
While both experimental and numerical data on the finite model reveal a bandgap between 1.990 and 3.355 kHz, the infinite model predicts a range between 2.021 and 3.427 kHz. The comparison implies a relative error of 1.5$\%$ in the lower frequency and of 2.3$\%$ in the upper frequency of the bandgap. We related this small difference to the limited number of unit cells and the presence of material damping when compared to the infinite model. 
The imaginary part of the wave-number retrieved experimentally (as shown in Figure. \ref{fig2-exp}d) validates the Bragg type bandgap mechanism. 
It should be mentioned that numerical calculations were initially performed based on the nominal values of the elastic parameters that the manufacturer provided in the data sheet for versatile plastic. The nominal value for Young's modulus was $E$ = 1.7 GPa, but after running the experiments, the experimental data revealed a bandgap which was shifted toward lower frequencies compared to the one expected on the basis of the infinite model and nominal values. Numerical testing determined that an  $E$ = 1 GPa results in a better match between the experimental and numerical results.  Throughout the paper, the numerical results are the ones obtained from the updated value for the Young's modulus. The discrepancy between the nominal value of the Young's modulus and the one of the 3D printed structure has also been reported in other studies \citep{Aguzzi-apl,beli2019}.

Once the dynamic properties of the Schwarz primitive cell are characterised, we may proceed with designing metastructures for propagation mitigation (metabarriers) and focusing (metalenses) purposes at a given frequency. Both devices are obtained by varying the mass distribution on the unit cells (either by adding or removing the mass) and in a target area on the plate. These variations yield a variation of the dispersion properties.

\section{Metabarriers for elastic wave mitigation}
\

\begin{figure}[h]
\centering
\includegraphics[trim={0 170 0 130},clip, keepaspectratio,scale=0.9]{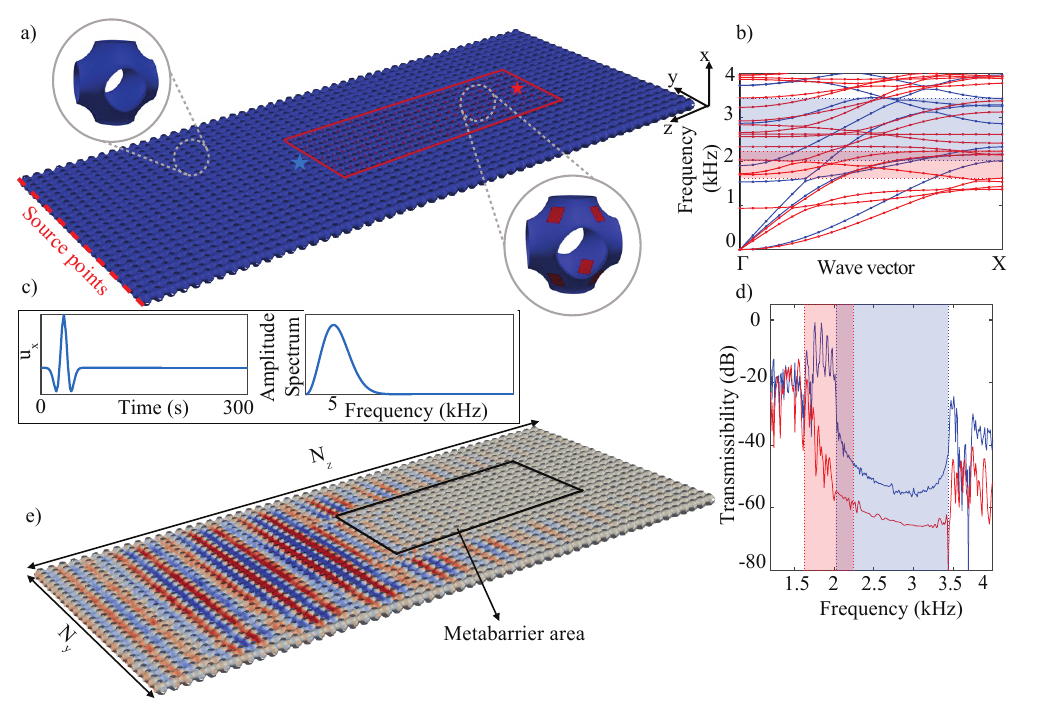}

\caption{(a) Configuration of the metabarrier (indicated by the red square) where the modified cell is the unit cell with the added mass. (b) Comparison between the dispersion relation of the standard unit cell (blue) and the modified unit cell (red) with the added mass density of 8000 kg/m$^3$. The bandgap range is marked by shaded areas in each case. (c)  Wavelet in time and frequency domain used to compute the transmissibility. (d)  Transmissibility measured before (blue) and at the end of the  metabarrier (red) at points marked by blue and red stars in panel (a). (e) Snapshot illustrating propagation of A$_0$ mode in the plate with an embedded metabarrier. }\label{fig3-mb_mass}
\end{figure}

The first step of the metabarrier design is to investigate the influence of increasing the mass of the unit cell on the lowest out-of-plane mode. A visual inspection of the bandgap opening/closing mode suggests to locate the mass where the inertia will be maximised, i.e., in the center of the unit cell face.
The comparison of dispersion relations and bandgap range between the standard unit cell and the one where a mass density of 8000 kg/m$^3$ is added to the unit cell \textcolor{black}{(see the insets in Figure.  \ref{fig3-mb_mass}a)} is shown in Figure.  \ref{fig3-mb_mass}b.  
While the standard unit cell has a bandgap between 2.021 and 3.427 kHz, the cell with the added mass has a bandgap between 1.624 and 2.163 kHz. Thus, we can observe adding the mass to the unit cell shifts the bandgap toward lower frequencies. We leverage this shift and the overlap between the two bandgaps to design the metabarrier.
To realize a metabarrier, a 2D plate configuration of $N_z$= 54 by $N_y$= 26 unit cells is designed (See Figure. \ref{fig3-mb_mass}a). Part of this plate is an interior area consisting of 10 by 24 unit cells along the $y$ and $z$ directions (as indicated with the red rectangle in Figure. \ref{fig3-mb_mass}a). Within this area rectangular portions of the unit cells, specified in red in the inset of Figure. \ref{fig3-mb_mass}a, have a higher mass density (8000 kg/m$^3$) than the standard cell.

Performance of the metabarrier is verified by exciting the out-of-plane mode ($x$ component) with a broadband Ricker wavelet centered at frequency of 5 kHz (See Figure. \ref{fig3-mb_mass}c). 
The transmissibility is computed before and after the metabarrier (the two points marked by stars in Figure. \ref{fig3-mb_mass}a) to evaluate its effect on the propagating wavefield. Figure. \ref{fig3-mb_mass}d shows that the transmissibility computed at the end of the metabarrier region unfolds a bandgap wider than the one of the standard unit cell. 
The starting frequency of this widened bandgap coincides with the opening frequency of the bandgap for the unit cell with the added mass (according to  Figure. \ref{fig3-mb_mass}b), and the closing frequency matches the one of the bandgap of the standard unit cell. 

Moreover, the bandgap computed at the position of the red star shows a stronger attenuation of the A$_0$ mode (65 dB as opposed to 52 dB). Thus, the metabarrier built by adding the mass to the unit cell leads to a wider and deeper attenuation.  
The performance of the metabarrier can be further validated by exciting a tapered sinusoid with the frequency of 1.850 kHz. Based on Figure. \ref{fig3-mb_mass}d, this frequency lies in a range where no propagation in the metabarrier region is expected. A snapshot of the results of such a simulation is shown in Figure. \ref{fig3-mb_mass}e. It can be clearly observed that at this frequency, no waves radiate in the metabarrier region  and that the waves are reflected backwards. This observation guarantees the possibility of protecting a given region on the plate at a certain frequency range.

\begin{figure}[h]
\centering
\includegraphics[trim={0 180 0 170},clip, keepaspectratio,scale=0.85]{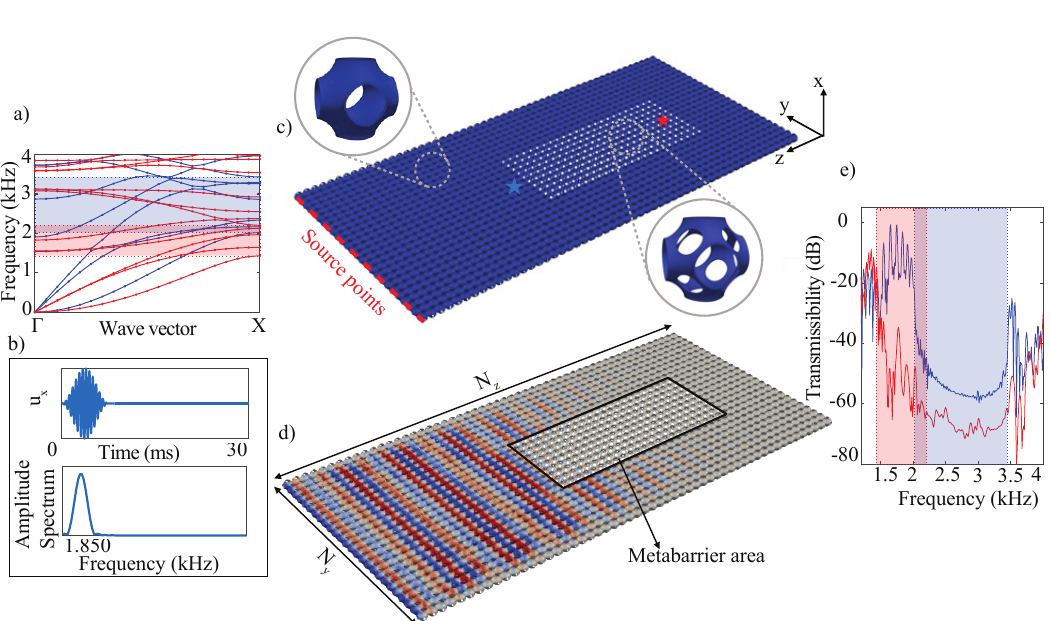}

\caption{(a) Comparison between the dispersion relation of the bare unit cell (blue) and the unit cell with the increased porosity of radius 5 mm (red). (b) Sinusoidal source wavelet in time and frequency domain used to validate the performance of the metabarrier. (c) Configuration of the metabarrier where the design parameter is the unit cell with the increased porosity. (d) Snapshot of the A$_0$ mode of the propagation in the plate with an embedded metabarrier. (e) Transmissibility measured before (blue) and at the end of the  metabarrier (red) at points marked by blue and red stars in panel (c).}\label{fig4-mb_hole}
\end{figure}

Next, we show the design of a metabarrier for the case, where we remove mass from the unit cell, i.e., by increasing the porosity.
To do so, spheres with a given radius were subtracted from the unit cell at eight positions and by gradually varying the radius of the sphere, i.e., the resulting increase in porosity, a \textcolor{black}{localized parametric analysis} was performed. 
An example of the unit cell with the removed mass (equivalently, increased porosity) is shown in the inset of Figure. \ref{fig4-mb_hole}c. A comparison between dispersion relations of the bare unit cell and the cell with the increased porosity where the radius of the hole is equal to 5 mm are shown in Figure. \ref{fig4-mb_hole}a. It turns out that increasing the porosity pushes the modes toward lower frequencies, which can be explained by the stronger role of compliance over the mass. This finding enables the design of a metabarrier based on the increased porosity guaranteeing a much lighter metamaterial compared to the added mass instance.
A metabarrier with the same size of the plate as the one for the case of the added mass is built and the configuration is illustrated in Figure. \ref{fig4-mb_hole}c.
The widening of the bandgap and the higher attenuation within the metabarrier are confirmed in Figure. \ref{fig4-mb_hole}e.
The results show the emergence of a bandgap between 1.431 and 3.427 kHz at the end of the metabarrier. Validation of the metabarrier performance is achieved by exciting the plate with a \textcolor{black}{Hann-windowed} sinusoidal at the frequency 1.850 kHz (the wavelet is shown in Figure. \ref{fig4-mb_hole}b). The results confirm the protection of the metabarrier region at this frequency through reflecting the waves back to the bare plate.  
This result proves the possibility of wave mitigation by further lightening the Schwarz primitive cells. 

\section{Metadevices for elastic wave focusing}
\

We have shown that by leveraging the variations in the dispersion relation after adding or removing mass to/from the unit cell, a metabarrier can be realized. Next, we leverage this effect further in order to focus the incoming waves. Rather than designating an area in the plate with a constant ramp in the unit cell's mass or porosity, we now follow a curvilinear profile defined by an equation.  In particular we are interested in the refractive index profile of a GRIN lens such as the Luneburg one \citep{luneburg1944mathematical}.
The feasibility of this lens has been previously proved in different types of resonant structures: rods attached to a thin plate \citep{andrea6}, octet lattice unit cells \citep{Aguzzi2022} in the kHz regime,  I-shaped metallic structure embedded in a substrate in GHz regime \citep{chen2015modified} and also sound focusing based on acoustic metamaterials  \citep{acousticmm-luneberg}.  However, in these studies the refractive index profile is satisfied by adding mass, increasing the height of rods, thus yielding a heavier zone compared to the surrounding host surface. 
Here, we tailor the bandgap by increasing the porosity. This is of particular interest, because the refractive index distribution is tuned by the size of the pore on the unit cell resulting in a lightweight meta area. 

\subsection{Design of a Luneburg lens with increased porosity}

\
The Luneburg lens is spherically symmetric and satisfies a parabolic refractive index profile ($n$), where $n$ must vary according to the Equation. \ref{eq:nLun} as a function of the radial distance:

\begin{equation}
    n = \sqrt{2-(\frac{r}{R})^2}.   
    \label{eq:nLun}
\end{equation}
Here, $r$ is the radius of each layer and $R$ is the radius of the full lens. In order to satisfy the requirement for the distribution of refractive index,  the velocity of the waves need to vary in each spherical layer within the lens region. The refractive index $n$ is defined as:

\begin{equation}
    n = \frac{v_0}{v}   
    \label{eq:ndef}
\end{equation}
where $v_0$ is the wave speed at a given frequency in the standard unit cell and $v$ is the wave speed at each layer of the metalens.

\begin{figure}[h!]
\centering
\includegraphics[trim={0 150 0 50},clip, keepaspectratio,scale=0.9]{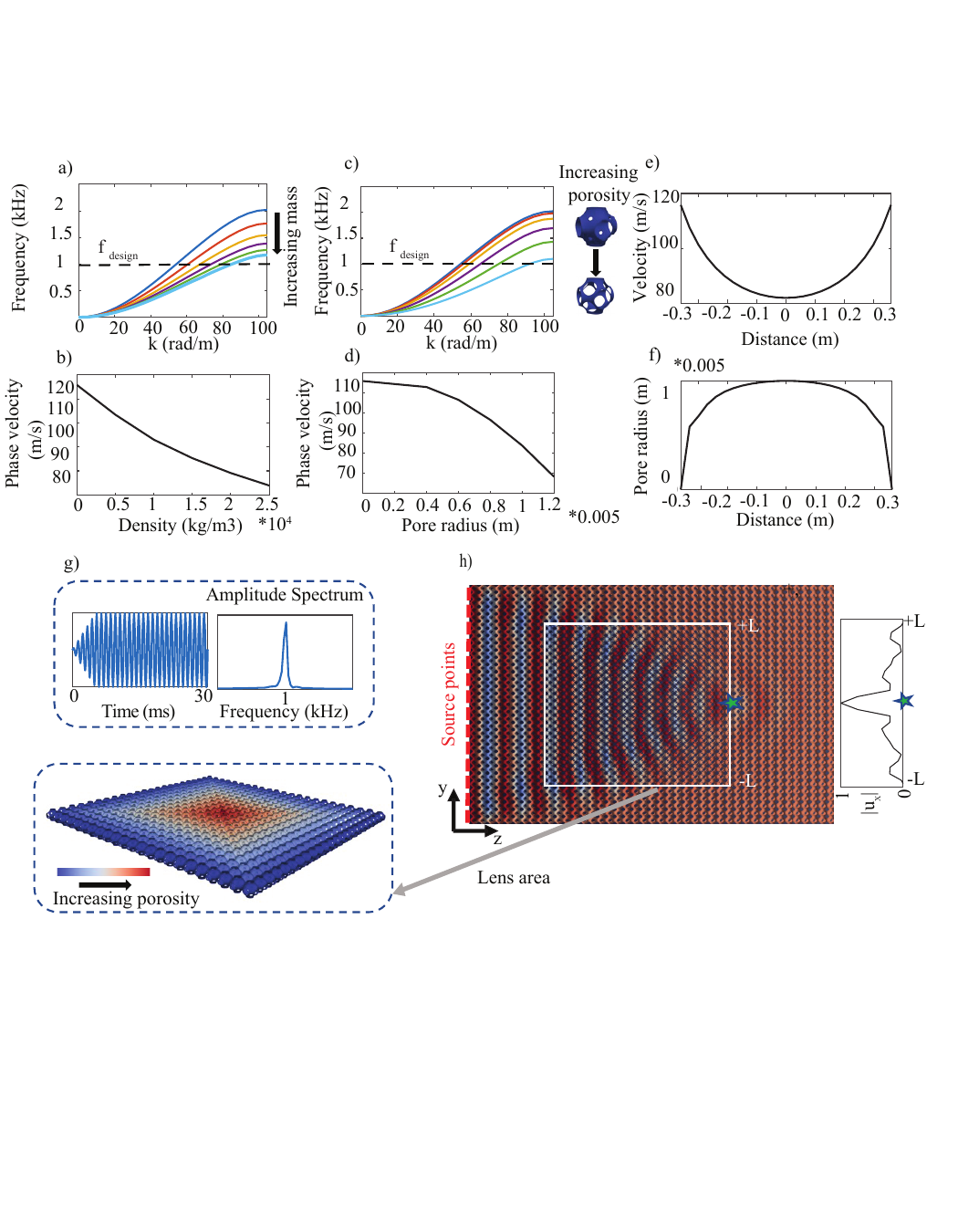}

\caption{(a) \textcolor{black}{localized parametric analysis} of the A$_0$ mode with the added mass. (b) Calculated phase speed based at the design frequency. (c) \textcolor{black}{localized parametric analysis} of the A$_0$ mode with the increased porosity. (d) Calculated phase speed at the design frequency. (e) Velocity profile for Luneburg lens as a function of the distance from the center of the lens. (f) Porosity profile for the Luneburg lens based on the Schwarz primitive cell.  (g) Source wavelet in the time and frequency domains. (h) Performance of the lens with square layers at frequency of 1 kHz and the focusing phenomenon. The colored inset shows the pore distribution inside the metalens area. }\label{fig5_lens_mass}
\end{figure}

Following the procedure outlined above for designing the Luneburg lens, we start by a \textcolor{black}{localized parametric analysis} of the unit cell with two different design parameters: namely as the added mass and the increased porosity, and the evolution of the dispersion relation of the $A_0$ mode in each case is tracked. \textcolor{black}{Values of added mass density are between 5000 and 25000 kg/m$^3$ and the pore radius for the case of increased porosity varies between 1 to 6 mm, which translates to a porosity range between 90\% and 95\%.} Phase speed is then calculated at the design frequency via $v=\omega/k$, where $\omega$ is the angular frequency and $k$ is the wavenumber. Figure. \ref{fig5_lens_mass}a and c show how the frequency range of the $A_0$ mode changes with increasing mass or porosity. 
We design the lens to be effective at the frequency of $f$= 1 kHz. Therefore, the phase speed at this frequency is calculated and its variations with the added mass are shown in Figure. \ref{fig5_lens_mass}b and increasing porosity in Figure. \ref{fig5_lens_mass}d. 
According to Equation. \ref{eq:nLun}, a Luneburg lens needs a contrast in the refraction index profile that goes from 1 (outermost layer) to 1.4 (central layer). This requires a velocity profile illustrated in Figure. \ref{fig5_lens_mass}e.

By knowing the required velocity profile for the Luneburg lens (from Figure. \ref{fig5_lens_mass}e) and the wave speed variations caused by added mass or porosity (from Figs. \ref{fig5_lens_mass}b and d), we can determine the necessary porosity (or added mass) for each layer of the lens to match the desired refractive index profile, as illustrated in Figure. \ref{fig5_lens_mass}f. It should be noted that a rectangular layer design, as depicted in the inset of Figure. \ref{fig5_lens_mass}h, is preferred due to the cubic cell shape and size, which naturally distributes porosity (or mass) minimising aberrations. Consequently, the radius $r$ in Equation. \ref{eq:nLun} can be translated into a distance from the center of the lens. 

The procedures for both increased porosity and added mass are identical. To simplify and avoid repetition, we present results for the more relevant \textcolor{black}{increased porosity} model. The numerical model of the Luneburg lens involves replicating unit cells in two directions within a plate. The lens region spans from $-L = -12*cell$ $size$ to $+L = +12*cell$ $size$, with a $cell$ $size$  of 3 cm. Value of the sphere radius for adding the porosity to each layer of the lens is determined by values reported in Figure.~\ref{fig5_lens_mass}f.
The out-of-plane mode is excited at one end of the plate with a sinusoidal signal at the design frequency (i.e., $f$ = 1 kHz), resulting in plane wave propagation in the $z$ direction. The source waveform in both the time and frequency domains is depicted in Figure. \ref{fig5_lens_mass}g. A snapshot of the propagating mode in Figure. \ref{fig5_lens_mass}h shows that waves inside the lens area (indicated by the white square) are slowed down and bent, so that they focus at the other end of the lens as expected, marked by the green star. The focusing spot is recovered by extracting the maximum out-of-plane displacement from $-L$ to $+L$, as shown in the right panel of the Figure.  \ref{fig5_lens_mass}h. Despite the fact that grading is based on a circular shape but applied to squared layers, we observe that focusing  at $f$= 1 kHz is achieved.

\subsection{Experimental validation of the Luneburg lens with increased porosity}
\begin{figure}[h!]
\centering
\includegraphics[trim={40 120 0 190},clip, keepaspectratio,scale=0.9]{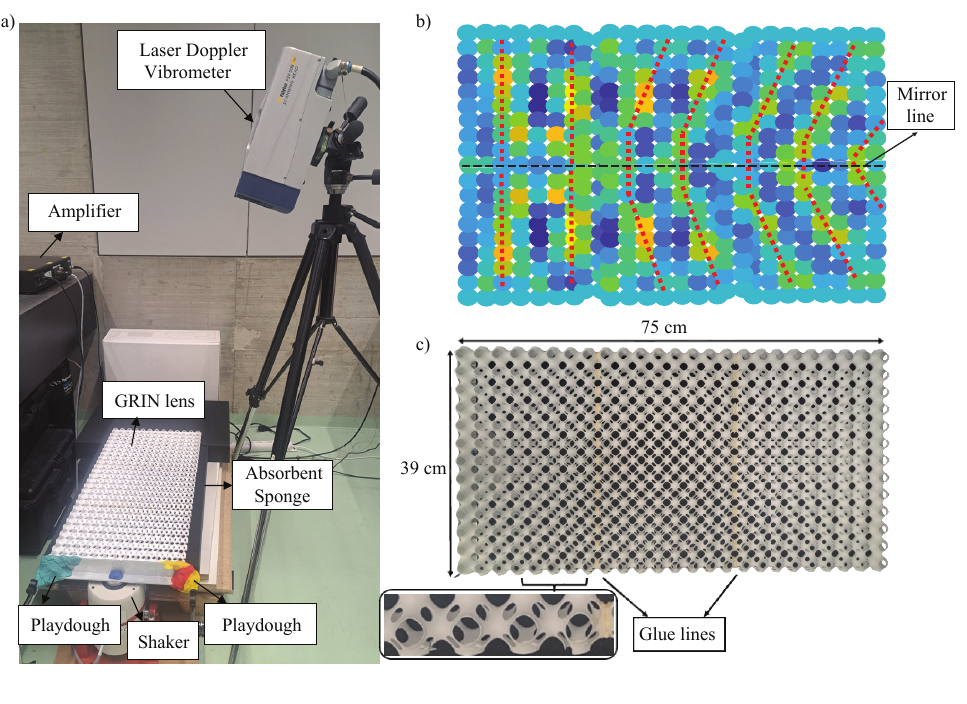}

\caption{(a) Experimental set-up for wave bending validation in the graded area of the Luneburg lens based on increased porosity. (b) Experimental measurement of the out-of-plane propagating mode at frequency of 1 kHz. The out-of-plane mode bending, marked by red dashed lines, is apparent during the propagation in the graded area. (c) 3D printed prototype of the Luneburg lens. The inset shows the cells on the border of the prototype, which are damaged through the 3D printing process and therefore the measured data at those points are excluded. }\label{fig6_lens_exp}
\end{figure}

\
We now show the experimental realisation of the Luneburg lens. 
The 3D printed prototype shown in Figure. \ref{fig6_lens_exp}c is the top half of the lens validated numerically (i.e., the top half of the white rectangle in Figure. \ref{fig5_lens_mass}h) and therefore is composed of 13 by 25 unit cells. Due to restrictions on the size of the prototype to be 3D printed, the design was divided into three sub samples, which were then 3D printed via the SLS method and glued together. The resulting structure is shown in Figure. \ref{fig6_lens_exp}c. A plate with dimensions of 39 by 3 by 0.4 cm is added to the bottom of the sample and is attached to a shaker to facilitate the excitation. The sample is positioned horizontally on an absorbent sponge in order to minimize the interactions with the ground. Additionally, some playdough has been added to the sides of the plate attached to the shaker to weaken the propagation of unwanted modes of said plate into the structure (See Figure.  \ref{fig6_lens_exp}a). Finally, the plate is excited with a sine wave of central frequency 1 kHz (i.e., the design frequency of the lens). An LDV is used to measure the resulting out-of-plane velocity of the propagating wavefield on a grid of 334 points. The resulting wavefield at $t$ = 34 ms is shown in Fig. \ref{fig6_lens_exp}b, where for the sake of a better visualization the grid points are mirrored horizontally in order to mimic the wavefield in a full size lens. The measurement points on the bottom lines of the structure are excluded because as shown in the inset of Figure. \ref{fig6_lens_exp}c, some boundary cells have come out of the 3D printing broken. Bending of the out-of-plane mode within the graded area clearly occurs in Figure. \ref{fig6_lens_exp}b, thus experimentally validating the analysed Luneburg lens design.

\section{Discussion}

\
Static properties of the Schwarz primitive cell are well known and research is ongoing to design porous and bone mimicking scaffolds based on such geometries for medical applications and bone implants. From a dynamic point of view, the performance of structures based on the Schwarz primitive cell for vibration and elastic wave guiding has been previously investigated  \citep{VIET, ma2023semi}. However, studies on the capacity of such structures in regard to wave mitigation and focusing in the context of metamaterials are missing. 
In this work, we showed that the Schwarz primitive cell can be used as the building block of metadevices for tailoring, mitigation and focusing of elastic waves. This validation paves the way for constructing further metadevices such as GRIN lenses with a hyperbolic secant profile for the refractive index profile, which are effective in a broader range of frequencies and not only at a single frequency.
Interestingly, we observed that both adding and removing the mass shift the A$_0$ mode toward lower frequencies which is an effective means to designing metabarriers and focusing devices. This behaviour is expected for the case of adding mass to the unit cell. 
However, it seems that by increasing porosity the compliance and lower stiffness have a greater effect than the resulting mass removal, thus, causing the shift of the out-of-plane mode toward lower frequencies.
The experimentally demonstrated case of increasing porosity to achieve wavefield focusing has further potential as it results in a much lighter structure \textcolor{black}{compared to previously-realized metadevices in the literature, which are on the basis of adding the mass to the structure or increasing the height of the rod resonators, resulting in a heavier metastructure.}. 
Devices based on the increased porosity are on one hand more feasible for 3D printing and therefore experimental validation. On the other hand, removing the mass changes the static properties of such cells and although the dynamic capacity of the cell based on the increased porosity is verified, research is needed to investigate the static properties of such lightweight structures. The material used in this study (i.e., versatile plastic) makes the structure fragile considering the delicacy of the unit cell, which limits application of metadevices for real world applications problems. In this regard, optimization of the pore shape and curvature as well as the use of an additional solid matrix and material optimization will be a necessary step to qualify metastructures based on the Schwarz primitive cell for a practical engineering usage. 

\section{Conclusion}
\
In this study, we retrieved the band diagram of a Schwarz primitive unit cell made of versatile plastic based on numerical investigations both in the time and frequency domain. We also validated the dispersion relation through laboratory experiments on the small scale printed prototypes using a 1D laser Doppler vibrometer. Both the numerical and experimental analyses revealed a bandgap of Bragg type that matches very well the bandgap expected from the infinite model. Next, the variations of the dispersion relation when adding mass or increasing porosity were leveraged in order to build metastructures such as metabarriers and GRIN metalenses. Both metabarriers show a wider and deeper attenuation performance, which indicates the capacity of the Schwarz primitive cell for further vibration mitigation. 
We also show that a Luneburg lens with a parabolic refractive index profile can be constructed for focusing the out-of-plane mode by both adding mass or porosity to the standard cell. Although the Luneburg lens is spherically symmetric and therefore is often designed based on circular layers of varying properties in the meta area, we show that the design still works with rectangular layers. 
We present a first experimental verification of wavefield focusing by increasing the porosity in an area of a plate constructed of Schwarz primitive unit cells. 
Our results encourage further application of bio-mimetic structures for tailoring and manipulating of elastic waves. 
Moreover, the proposed unit cell is scalable and therefore tunable to perform at a given frequency of interest leading to novel functionalities in elastic wave propagation including energy harvesting.

\section{Acknowledgements}
\
A.H.N and A.C were supported by the H2020 FETOpen project BOHEME under Grant Agreement No. 863179, and A.C, H.R.T and B.Z acknowledge funding from the H2020 FET-proactive project MetaVEH, European Union under the Grant Agreement No. 952039. This work was supported by a grant from the Swiss National Supercomputing Centre (CSCS) under project ID s1185.
Laboratory Equipment was provided to A.C. through the SNSF R'Equip grant 205418. 
\bibliography{references}

\begin{thebibliography}{47}
\providecommand{\natexlab}[1]{#1}
\providecommand{\url}[1]{\texttt{#1}}
\expandafter\ifx\csname urlstyle\endcsname\relax
  \providecommand{\doi}[1]{doi: #1}\else
  \providecommand{\doi}{doi: \begingroup \urlstyle{rm}\Url}\fi

\bibitem[Tang et~al.(2017)Tang, Ren, Meng, Xin, Huang, Chen, Zhang, and Lu]{acoustic-absorption}
Yufan Tang, Shuwei Ren, Han Meng, Fengxian Xin, Lixi Huang, Tianning Chen, Chuanzeng Zhang, and Tian~Jian Lu.
\newblock Hybrid acoustic metamaterial as super absorber for broadband low-frequency sound.
\newblock \emph{Scientific Reports}, 2017.
\newblock \doi{10.1038/srep43340}.
\newblock URL \url{https://doi.org/10.1038/srep43340}.

\bibitem[Park et~al.(2020)Park, Lee, and Rho]{park-guiding}
Jeonghoon Park, Dongwoo Lee, and Junsuk Rho.
\newblock Recent advances in non-traditional elastic wave manipulation by macroscopic artificial structures.
\newblock \emph{Applied Sciences}, 10\penalty0 (2), 2020.
\newblock ISSN 2076-3417.
\newblock \doi{10.3390/app10020547}.
\newblock URL \url{https://www.mdpi.com/2076-3417/10/2/547}.

\bibitem[Chen et~al.(2014)Chen, Guo, Yang, and Cheng]{CHEN-harvesting-review}
Zhongsheng Chen, Bin Guo, Yongmin Yang, and Congcong Cheng.
\newblock Metamaterials-based enhanced energy harvesting: A review.
\newblock \emph{Physica B: Condensed Matter}, 438:\penalty0 1--8, 2014.
\newblock ISSN 0921-4526.
\newblock \doi{https://doi.org/10.1016/j.physb.2013.12.040}.
\newblock URL \url{https://www.sciencedirect.com/science/article/pii/S092145261300834X}.

\bibitem[Zhao et~al.(2022)Zhao, Thomsen, {De Ponti}, Riva, {Van Damme}, Bergamini, Chatzi, and Colombi]{Zhao-harvesting}
Bao Zhao, Henrik~R. Thomsen, Jacopo~M. {De Ponti}, Emanuele Riva, Bart {Van Damme}, Andrea Bergamini, Eleni Chatzi, and Andrea Colombi.
\newblock A graded metamaterial for broadband and high-capability piezoelectric energy harvesting.
\newblock \emph{Energy Conversion and Management}, 269:\penalty0 116056, 2022.
\newblock ISSN 0196-8904.
\newblock \doi{https://doi.org/10.1016/j.enconman.2022.116056}.
\newblock URL \url{https://www.sciencedirect.com/science/article/pii/S0196890422008445}.

\bibitem[Dubček et~al.(2023)Dubček, Moreno-Garcia, Haag, Omidvar, Thomsen, Becker, Gebraad, Bärlocher, Andersson, Huber, van Manen, Villanueva, Robertsson, and Serra-Garcia]{dub2023binary}
Tena Dubček, Daniel Moreno-Garcia, Thomas Haag, Parisa Omidvar, Henrik~R. Thomsen, Theodor~S. Becker, Lars Gebraad, Christoph Bärlocher, Fredrik Andersson, Sebastian~D. Huber, Dirk-Jan van Manen, Luis~Guillermo Villanueva, Johan O.~A. Robertsson, and Marc Serra-Garcia.
\newblock Binary classification of spoken words with passive phononic metamaterials, 2023.

\bibitem[Miniaci et~al.(2016)Miniaci, Krushynska, Bosia, and Pugno]{Miniaci-seismic-shielding}
Marco Miniaci, Anastasiia Krushynska, Federico Bosia, and Nicola~M Pugno.
\newblock Large scale mechanical metamaterials as seismic shields.
\newblock \emph{New Journal of Physics}, 18\penalty0 (8):\penalty0 083041, aug 2016.
\newblock \doi{10.1088/1367-2630/18/8/083041}.
\newblock URL \url{https://dx.doi.org/10.1088/1367-2630/18/8/083041}.

\bibitem[Zhao et~al.(2023)Zhao, Thomsen, Pu, Fang, Lai, Van~Damme, Bergamini, Chatzi, and Colombi]{zhao2023nonlinear}
Bao Zhao, Henrik~R Thomsen, Xingbo Pu, Shitong Fang, Zhihui Lai, Bart Van~Damme, Andrea Bergamini, Eleni Chatzi, and Andrea Colombi.
\newblock A nonlinear damped metamaterial: Wideband attenuation with nonlinear bandgap and modal dissipation.
\newblock \emph{arXiv preprint arXiv:2307.14165}, 2023.

\bibitem[Chondrogiannis et~al.(2022)Chondrogiannis, Colombi, Dertimanis, and Chatzi]{kyriakosPhysRevApplied}
Kyriakos~Alexandros Chondrogiannis, Andrea Colombi, Vasilis Dertimanis, and Eleni Chatzi.
\newblock Computational verification and experimental validation of the vibration-attenuation properties of a geometrically nonlinear metamaterial design.
\newblock \emph{Phys. Rev. Appl.}, 17:\penalty0 054023, May 2022.
\newblock \doi{10.1103/PhysRevApplied.17.054023}.
\newblock URL \url{https://link.aps.org/doi/10.1103/PhysRevApplied.17.054023}.

\bibitem[Brillouin(2003)]{brillouin}
L.~Brillouin.
\newblock \emph{Wave Propagation in Periodic Structures: Electric Filters and Crystal Lattices}.
\newblock Dover phoenix editions. Dover Publications, 2003.
\newblock ISBN 9780486495569.
\newblock URL \url{https://books.google.ch/books?id=m2WmGiU5nUwC}.

\bibitem[Kittel(2004)]{kittel2004introduction}
C.~Kittel.
\newblock \emph{Introduction to Solid State Physics}.
\newblock Wiley, 2004.
\newblock ISBN 9780471415268.
\newblock URL \url{https://books.google.ch/books?id=kym4QgAACAAJ}.

\bibitem[Colombi et~al.(2016)Colombi, Roux, Guenneau, Gueguen, and Craster]{Colombi2016forests}
Andrea Colombi, Philippe Roux, Sebastien Guenneau, Philippe Gueguen, and Richard~V Craster.
\newblock Forests as a natural seismic metamaterial: {R}ayleigh wave bandgaps induced by local resonances.
\newblock \emph{Scientific Reports}, 6\penalty0 (1):\penalty0 1--7, 2016.

\bibitem[Zhou et~al.(2012)Zhou, Liu, and Hu]{ZHOU2012041001}
Xiaoming Zhou, Xiaoning Liu, and Gengkai Hu.
\newblock Elastic metamaterials with local resonances: an overview.
\newblock \emph{Theoretical and Applied Mechanics Letters}, 2\penalty0 (4):\penalty0 041001, 2012.
\newblock ISSN 2095-0349.
\newblock \doi{https://doi.org/10.1063/2.1204101}.
\newblock URL \url{https://www.sciencedirect.com/science/article/pii/S2095034915301586}.

\bibitem[Cenedese et~al.(2021)Cenedese, Belloni, and Braghin]{cenedese-coale}
Mattia Cenedese, Edoardo Belloni, and Francesco Braghin.
\newblock {Interaction of Bragg scattering bandgaps and local resonators in mono-coupled periodic structures}.
\newblock \emph{Journal of Applied Physics}, 129\penalty0 (12), 03 2021.
\newblock ISSN 0021-8979.
\newblock \doi{10.1063/5.0038438}.
\newblock URL \url{https://doi.org/10.1063/5.0038438}.
\newblock 124501.

\bibitem[Krushynska et~al.(2017)Krushynska, Miniaci, Bosia, and Pugno]{KRUSHYNSKA201730}
A.O. Krushynska, M.~Miniaci, F.~Bosia, and N.M. Pugno.
\newblock Coupling local resonance with bragg band gaps in single-phase mechanical metamaterials.
\newblock \emph{Extreme Mechanics Letters}, 12:\penalty0 30--36, 2017.
\newblock ISSN 2352-4316.
\newblock \doi{https://doi.org/10.1016/j.eml.2016.10.004}.
\newblock URL \url{https://www.sciencedirect.com/science/article/pii/S2352431616300542}.
\newblock Frontiers in Mechanical Metamaterials.

\bibitem[Lee and Iizuka(2019)]{lee-resonance-bragg}
Taehwa Lee and Hideo Iizuka.
\newblock Bragg scattering based acoustic topological transition controlled by local resonance.
\newblock \emph{Phys. Rev. B}, 99:\penalty0 064305, Feb 2019.
\newblock \doi{10.1103/PhysRevB.99.064305}.
\newblock URL \url{https://link.aps.org/doi/10.1103/PhysRevB.99.064305}.

\bibitem[Aguzzi et~al.(2022{\natexlab{a}})Aguzzi, Kanellopoulos, Wiltshaw, Craster, Chatzi, and Colombi]{Aguzzi2022}
Giulia Aguzzi, Constantinos Kanellopoulos, Richard Wiltshaw, Richard~V Craster, Eleni~N Chatzi, and Andrea Colombi.
\newblock Octet lattice-based plate for elastic wave control.
\newblock \emph{Scientific Reports}, 12\penalty0 (1):\penalty0 1--14, 2022{\natexlab{a}}.

\bibitem[Rupin et~al.(2014)Rupin, Lemoult, Lerosey, and Roux]{rupin2014experimental}
Matthieu Rupin, Fabrice Lemoult, Geoffroy Lerosey, and Philippe Roux.
\newblock Experimental demonstration of ordered and disordered multiresonant metamaterials for lamb waves.
\newblock \emph{Physical review letters}, 112\penalty0 (23):\penalty0 234301, 2014.

\bibitem[Datta et~al.(2022)Datta, Tamburrino, and Udpa]{Datta-microwave}
Srijan Datta, Antonello Tamburrino, and Lalita Udpa.
\newblock Gradient index metasurface lens for microwave imaging.
\newblock \emph{Sensors}, 22\penalty0 (21), 2022.
\newblock ISSN 1424-8220.
\newblock \doi{10.3390/s22218319}.
\newblock URL \url{https://www.mdpi.com/1424-8220/22/21/8319}.

\bibitem[Fuentes-Dominguez et~al.(2021)Fuentes-Dominguez, Yao, Colombi, Dryburgh, Pieris, Jackson-Crisp, Colquitt, Clare, Smith, and Clark]{SAW}
Rafael Fuentes-Dominguez, Mengting Yao, Andrea Colombi, Paul Dryburgh, Don Pieris, Alexander Jackson-Crisp, Daniel Colquitt, Adam Clare, Richard~J. Smith, and Matt Clark.
\newblock Design of a resonant luneburg lens for surface acoustic waves.
\newblock \emph{Ultrasonics}, 111:\penalty0 106306, 2021.
\newblock ISSN 0041-624X.
\newblock \doi{https://doi.org/10.1016/j.ultras.2020.106306}.
\newblock URL \url{https://www.sciencedirect.com/science/article/pii/S0041624X20302419}.

\bibitem[Climente et~al.(2014)Climente, Torrent, and S{\'a}nchez-Dehesa]{climente-gradient}
Alfonso Climente, Daniel Torrent, and Jos{\'e} S{\'a}nchez-Dehesa.
\newblock Gradient index lenses for flexural waves based on thickness variations.
\newblock \emph{Applied Physics Letters}, 105\penalty0 (6):\penalty0 064101, 2014.

\bibitem[Schwarz(1972)]{schwarz1972gesammelte}
Hermann~Amandus Schwarz.
\newblock \emph{Gesammelte mathematische abhandlungen}, volume 260.
\newblock American Mathematical Soc., 1972.

\bibitem[Montazerian et~al.(2017)Montazerian, Davoodi, Asadi-Eydivand, Kadkhodapour, and Solati-Hashjin]{MONTAZERIAN201798}
H~Montazerian, E~Davoodi, M~Asadi-Eydivand, J~Kadkhodapour, and M~Solati-Hashjin.
\newblock Porous scaffold internal architecture design based on minimal surfaces: A compromise between permeability and elastic properties.
\newblock \emph{Materials \& Design}, 126:\penalty0 98--114, 2017.
\newblock ISSN 0264-1275.
\newblock \doi{https://doi.org/10.1016/j.matdes.2017.04.009}.
\newblock URL \url{https://www.sciencedirect.com/science/article/pii/S0264127517303593}.

\bibitem[Vijayavenkataraman et~al.(2018)Vijayavenkataraman, Zhang, Zhang, Hsi~Fuh, and Lu]{biomimetic}
Sanjairaj Vijayavenkataraman, Lei Zhang, Shuo Zhang, Jerry~Ying Hsi~Fuh, and Wen~Feng Lu.
\newblock Triply periodic minimal surfaces sheet scaffolds for tissue engineering applications: An optimization approach toward biomimetic scaffold design.
\newblock \emph{ACS Appl. Bio Mater}, 1, 2018.
\newblock URL \url{https://doi.org/10.1021/acsabm.8b00052}.

\bibitem[Pan et~al.(2020)Pan, Han, and Lu]{pan2020}
Chen Pan, Yafeng Han, and Jiping Lu.
\newblock Design and optimization of lattice structures: A review.
\newblock \emph{Applied Sciences}, 10\penalty0 (18), 2020.
\newblock ISSN 2076-3417.
\newblock \doi{10.3390/app10186374}.
\newblock URL \url{https://www.mdpi.com/2076-3417/10/18/6374}.

\bibitem[Sokollu et~al.(2022)Sokollu, Gulcan, and Konukseven]{SOKOLLU2022103199}
Baris Sokollu, Orhan Gulcan, and Erhan~Ilhan Konukseven.
\newblock Mechanical properties comparison of strut-based and triply periodic minimal surface lattice structures produced by electron beam melting.
\newblock \emph{Additive Manufacturing}, 60:\penalty0 103199, 2022.
\newblock ISSN 2214-8604.
\newblock \doi{https://doi.org/10.1016/j.addma.2022.103199}.
\newblock URL \url{https://www.sciencedirect.com/science/article/pii/S2214860422005887}.

\bibitem[Viet and Zaki(2021)]{VIET}
N.V Viet and W~Zaki.
\newblock Free vibration and buckling characteristics of functionally graded beams with triply periodic minimal surface architecture.
\newblock \emph{Composite Structures}, 274:\penalty0 114342, 2021.
\newblock ISSN 0263-8223.
\newblock \doi{https://doi.org/10.1016/j.compstruct.2021.114342}.
\newblock URL \url{https://www.sciencedirect.com/science/article/pii/S0263822321008047}.

\bibitem[Guo et~al.(2022)Guo, Rosa, Gupta, Dolan, Fields, Valdevit, and Ruzzene]{guo2022minimal}
Yuning Guo, Matheus Inguaggiato~Nora Rosa, Mohit Gupta, Benjamin~Emerson Dolan, Brandon Fields, Lorenzo Valdevit, and Massimo Ruzzene.
\newblock Minimal surface-based materials for topological elastic wave guiding.
\newblock \emph{Advanced Functional Materials}, 32\penalty0 (30):\penalty0 2204122, 2022.

\bibitem[Ma and Cheng(2023)]{ma2023semi}
Yongbin Ma and Qingfeng Cheng.
\newblock Semi-analytical solutions for wave propagation of periodically repetitive schwarz primitive triply periodic minimal surface based structure embedded with acoustic black hole resonators.
\newblock \emph{International Journal of Acoustics \& Vibration}, 28\penalty0 (3), 2023.

\bibitem[Tarcisio et~al.(2023)Tarcisio, Jin-You, K., and Dong-Wook]{silva2023}
Silva Tarcisio, Lu~Jin-You, Abu Al-Rub~Rashid K., and Lee Dong-Wook.
\newblock Investigation on tailoring the width and central frequency of bandgaps of tpms structures.
\newblock \emph{International Journal of Mechanics and Materials in Design}, 2023.
\newblock ISSN 1573-8841.
\newblock \doi{10.1007/s10999-023-09677-2}.
\newblock URL \url{https://doi.org/10.1007/s10999-023-09677-2}.

\bibitem[Jin-You et~al.(2023)Jin-You, Tarcisio, Fatima, K, and Dong-Wook]{Lu2023}
Lu~Jin-You, Silva Tarcisio, Alzaabi Fatima, Abu Al-Rub~Rashid K, and Lee Dong-Wook.
\newblock Insights into acoustic properties of seven selected triply periodic minimal surfaces-based structures: A numerical study.
\newblock \emph{Journal of Low Frequency Noise, Vibration and Active Control}, 43, 2023.
\newblock \doi{doi: 10.1177/14613484231190986}.
\newblock URL \url{https://doi.org/10.1177/14613484231190986}.

\bibitem[Elmadih et~al.(2019)Elmadih, Syam, Maskery, Chronopoulos, and Leach]{ELMADIH}
Waiel Elmadih, Wahyudin~P. Syam, Ian Maskery, Dimitrios Chronopoulos, and Richard Leach.
\newblock Mechanical vibration bandgaps in surface-based lattices.
\newblock \emph{Additive Manufacturing}, 25:\penalty0 421--429, 2019.
\newblock ISSN 2214-8604.
\newblock \doi{https://doi.org/10.1016/j.addma.2018.11.011}.
\newblock URL \url{https://www.sciencedirect.com/science/article/pii/S2214860418302781}.

\bibitem[Lai et~al.(2007)Lai, Kulak, Law, Zhang, Meldrum, and Riley]{Min-seaurchin}
Min Lai, Alex~N. Kulak, Daniel Law, Zhibing Zhang, Fiona~C. Meldrum, and D.~Jason Riley.
\newblock Profiting from nature: macroporous copper with superior mechanical properties.
\newblock \emph{Chem. Commun.}, pages 3547--3549, 2007.
\newblock \doi{10.1039/B707469G}.
\newblock URL \url{http://dx.doi.org/10.1039/B707469G}.

\bibitem[Al-Ketan and Abu Al-Rub(2019)]{Alketan-review}
Oraib Al-Ketan and Rashid~K. Abu Al-Rub.
\newblock Multifunctional mechanical metamaterials based on triply periodic minimal surface lattices.
\newblock \emph{Advanced Engineering Materials}, 21\penalty0 (10):\penalty0 1900524, 2019.
\newblock \doi{https://doi.org/10.1002/adem.201900524}.
\newblock URL \url{https://onlinelibrary.wiley.com/doi/abs/10.1002/adem.201900524}.

\bibitem[Gandy and Klinowski(2000)]{GANDY2000579}
Paul~J.F. Gandy and Jacek Klinowski.
\newblock Exact computation of the triply periodic schwarz p minimal surface.
\newblock \emph{Chemical Physics Letters}, 322\penalty0 (6):\penalty0 579--586, 2000.
\newblock ISSN 0009-2614.
\newblock \doi{https://doi.org/10.1016/S0009-2614(00)00453-X}.
\newblock URL \url{https://www.sciencedirect.com/science/article/pii/S000926140000453X}.

\bibitem[Abueidda et~al.(2018)Abueidda, Jasiuk, and Sobh]{ABUEIDDA201820}
Diab~W. Abueidda, Iwona Jasiuk, and Nahil~A. Sobh.
\newblock Acoustic band gaps and elastic stiffness of pmma cellular solids based on triply periodic minimal surfaces.
\newblock \emph{Materials \& Design}, 145:\penalty0 20--27, 2018.
\newblock ISSN 0264-1275.
\newblock \doi{https://doi.org/10.1016/j.matdes.2018.02.032}.
\newblock URL \url{https://www.sciencedirect.com/science/article/pii/S0264127518301163}.

\bibitem[Royer and Dieulesaint(1980)]{Royer1980ElasticWI}
Daniel Royer and Eug{\`e}ne Dieulesaint.
\newblock Elastic waves in solids.
\newblock 1980.

\bibitem[v.~5.4.()]{comsol_manual}
COMSOL~Multiphysics® v.~5.4.
\newblock \emph{Structural Mechanics Module User’s Guide}.

\bibitem[Alleyne and Cawley(1991)]{2d_fft}
D.~Alleyne and P.~Cawley.
\newblock {A two-dimensional Fourier transform method for the measurement of propagating multimode signals}.
\newblock \emph{The Journal of the Acoustical Society of America}, 89\penalty0 (3):\penalty0 1159--1168, 03 1991.
\newblock ISSN 0001-4966.
\newblock \doi{10.1121/1.400530}.
\newblock URL \url{https://doi.org/10.1121/1.400530}.

\bibitem[Bloch(1929)]{bloch1929quantenmechanik}
Felix Bloch.
\newblock {\"U}ber die quantenmechanik der elektronen in kristallgittern.
\newblock \emph{Zeitschrift f{\"u}r physik}, 52\penalty0 (7-8):\penalty0 555--600, 1929.

\bibitem[Komatitsch and Tromp(1999)]{Komatitsch1999}
Dimitri Komatitsch and Jeroen Tromp.
\newblock {Introduction to the spectral element method for three-dimensional seismic wave propagation}.
\newblock \emph{Geophysical Journal International}, 139\penalty0 (3):\penalty0 806--822, 12 1999.
\newblock ISSN 0956-540X.
\newblock \doi{10.1046/j.1365-246x.1999.00967.x}.
\newblock URL \url{https://doi.org/10.1046/j.1365-246x.1999.00967.x}.

\bibitem[Van~Damme and Zemp(2018)]{VanDamme-iwc}
Bart Van~Damme and Armin Zemp.
\newblock Measuring dispersion curves for bending waves in beams: A comparison of spatial fourier transform and inhomogeneous wave correlation.
\newblock \emph{Acta Acustica united with Acustica}, 104:\penalty0 228--234, 03 2018.
\newblock \doi{10.3813/AAA.919164}.

\bibitem[Aguzzi et~al.(2022{\natexlab{b}})Aguzzi, Thomsen, Hejazi~Nooghabi, Wiltshaw, Craster, Chatzi, and Colombi]{Aguzzi-apl}
Giulia Aguzzi, Henrik~R. Thomsen, Aida Hejazi~Nooghabi, Richard Wiltshaw, Richard~V. Craster, Eleni~N. Chatzi, and Andrea Colombi.
\newblock {Architected frames for elastic wave attenuation: Experimental validation and local tuning via affine transformation}.
\newblock \emph{Applied Physics Letters}, 121\penalty0 (20), 11 2022{\natexlab{b}}.
\newblock ISSN 0003-6951.
\newblock \doi{10.1063/5.0119903}.
\newblock URL \url{https://doi.org/10.1063/5.0119903}.
\newblock 201702.

\bibitem[Beli et~al.(2019)Beli, Fabro, Ruzzene, and Arruda]{beli2019}
Danilo Beli, Adriano Fabro, Massimo Ruzzene, and J.R.F. Arruda.
\newblock Wave attenuation and trapping in 3d printed cantilever-in-mass metamaterials with spatially correlated variability.
\newblock \emph{Scientific Reports}, 9, 04 2019.
\newblock \doi{10.1038/s41598-019-41999-0}.

\bibitem[Luneburg et~al.(1944)Luneburg, Herzberger, and School]{luneburg1944mathematical}
R.K. Luneburg, M.~Herzberger, and Brown University.~Graduage School.
\newblock \emph{Mathematical Theory of Optics: Supplementary Notes by M. Herzberger}.
\newblock Number Bd. 1 in Mathematical Theory of Optics: Supplementary Notes by M. Herzberger. 1944.
\newblock URL \url{https://books.google.ch/books?id=DAsJAQAAMAAJ}.

\bibitem[Colombi(2016)]{andrea6}
A.~Colombi.
\newblock Resonant metalenses for flexural waves.
\newblock \emph{J.\ Acoust.\ Soc.\ Am.}, 140\penalty0 (5):\penalty0 EL423, 2016.

\bibitem[Chen et~al.(2015)Chen, Cheng, Huang, Dai, Lu, Zhao, Ma, Jiang, and Cui]{chen2015modified}
Haibing Chen, Qiang Cheng, Aihua Huang, Junyan Dai, Huiying Lu, Jie Zhao, Huifeng Ma, Weixiang Jiang, and Tiejun Cui.
\newblock Modified luneburg lens based on metamaterials.
\newblock \emph{International Journal of Antennas and Propagation}, 2015, 2015.

\bibitem[Yuan et~al.(2022)Yuan, Liu, Long, Cheng, and Liu]{acousticmm-luneberg}
Baoguo Yuan, Jiyu Liu, Houyou Long, Ying Cheng, and Xiaojun Liu.
\newblock {Sound focusing by a broadband acoustic Luneburg lens}.
\newblock \emph{The Journal of the Acoustical Society of America}, 151\penalty0 (3):\penalty0 2238--2244, 03 2022.
\newblock ISSN 0001-4966.
\newblock \doi{10.1121/10.0009909}.
\newblock URL \url{https://doi.org/10.1121/10.0009909}.

\end{thebibliography}

\bibliographystyle{unsrtnat}
\end{document}